\documentclass{emulateapj}

\slugcomment{Draft: \today\ ; Accepted for publication in ApJL}

\shorttitle{High z candidates in submm surveys}
\shortauthors{Pope et al.}

\begin{document}

\title{Searching for the highest redshift sources in 250--500$\,\mu$m submillimeter surveys}

\author{Alexandra Pope\altaffilmark{1,2} \&
Ranga-Ram Chary\altaffilmark{3}
}

\altaffiltext{1}{National Optical Astronomy Observatory, 950 N. Cherry Ave., Tucson, AZ, 85719, USA}
\altaffiltext{2}{{\it Spitzer} Fellow; pope@noao.edu}
\altaffiltext{3}{Spitzer Science Center, California Institute of Technology, Pasadena, CA, 91125, USA}

\begin{abstract}

We explore a technique for identifying the highest redshift ($z>4$) sources in {\it Herschel}/SPIRE and BLAST submillimeter surveys by localizing the position of the far-infrared dust peak. Just as {\it Spitzer}/IRAC was used to identify stellar `bump' sources, the far-IR peak is also a redshift indicator; although, the latter also depends on the average dust temperature. We demonstrate the wide range of allowable redshifts for a reasonable range of dust temperatures and show that it is impossible to constraint the redshift of individual objects using solely the position of the far-IR peak. 
By fitting spectral energy distribution models to simulated {\it Herschel}/SPIRE photometry we show the utility of radio and/or far-infrared data in breaking this degeneracy. 
With prior knowledge of the dust temperature distribution it is possible to obtain statistical samples of high redshift submillimeter galaxy candidates. 
We apply this technique to the BLAST survey of ECDFS to constrain the number of dusty galaxies at $z>4$. 
We find $8\pm2$ galaxies with flux density ratios of $S_{500}>S_{350}$; this sets an upper limit of $17\pm4\,$deg$^{-2}$ if we assume all are at $z>4$.
This is $<35\,$\% of all 500$\,\mu$m-selected galaxies down to S$_{500}>45\,$mJy ($L_{\rm{IR}}>2\times10^{13}L_{\odot}$ for $z>4$).
Modeling with conventional temperature and redshift distributions estimates the percentage of these 500$\,\mu$m peak galaxies at $z>4$ to be between 10--85\%. 
Our results are consistent with other estimates of the number density of very high redshift submillimeter galaxies and follows the decline in the star formation rate density at $z>4$.

\end{abstract}

\keywords{galaxies: evolution --- galaxies: starburst --- galaxies: high-redshift --- infrared: galaxies --- submillimeter --- techniques: photometric}

\section{Introduction}
\label{sec:intro}

The prevalence of dusty, infrared luminous galaxies at early times tests our theories about early galaxy formation and the production of metals. 
Semi-analytic models predict very small numbers of submillimeter galaxies (SMGs) at extreme redshifts ($z>4$, e.g.~Baugh et al.~2005), and the best observational constraints on the redshift distribution of SMGs are consistent with this (Chapman et al.~2005; Swinbank et al.~2009). 
Much attention has been given to recent studies which have spectroscopically confirmed a small number of SMGs at $z>4$ (Capak et al.~2008; Coppin et al.~2009; Daddi et al.~2009ab; Knudsen et al.~2010); however, these sources are still only a small fraction of the submillimeter population. Small number statistics and cosmic variance in current submillimeter surveys prohibit an accurate measure of the space density of the most distant SMGs. 

The question of the prominence of the high redshift tail of dusty SMGs can be addressed with imminent submillimeter wide area surveys such as those with the {\it Herschel Space Observatory} (Pilbratt 2001), the SCUBA-2 camera on the JCMT (Holland et al.~2006), and the AzTEC camera on the Large Millimeter Telescope (LMT, Wilson et al.~2008). 
The {\it Herschel Space Observatory} is currently collecting data from 70--500$\,\mu$m over wide survey fields. With an aperture of 3.5$\,$m, {\it Herschel} is only sensitive to ultra-luminous infrared galaxies (ULIRGs, L$_{IR}>10^{12}L_{\odot}$) above $z>2$ at 250--500$\,\mu$m before it reaches the confusion limit. One advantage of SPIRE surveys is the simultaneous observations at three wavelengths; 250, 350 and 500$\,\mu$m, which samples the peak of the far-IR spectral energy distribution (SED) due to dust emission in high redshift galaxies. With prior
knowledge of the dust properties which describe the shape of the SED, the position of the far-IR peak is sensitive to the redshift of the galaxy. 

In this paper, we discuss the power and limitations of using the far-IR peak as a redshift indicator in submillimeter surveys. We apply this technique to the current BLAST survey of ECDFS and place an upper limit on the number density of 500$\,\mu$m-selected SMGs at $z>4$.

Throughout this paper we assume a standard cosmology with $H_{0}=71\,\rm{km}\,\rm{s}^{-1}\,\rm{Mpc}^{-1}$, $\Omega_{\rm{M}}=0.27$ and $\Omega_{\Lambda}=0.73$.

\section{Far-IR `bump' technique}

The 1.6\,$\mu$m bump in the near-infrared SED of galaxies arises due to a minimum in the H$^{-}$ opacity
in the spectra of cool stars. This bump
has been used in extragalactic {\it Spitzer}/IRAC surveys to identify sources in a specific redshift range (e.g.~Wright et al. 1994, 
Sawicki 2002, Farrah et al.~2008). A similar, but factor of $\sim2-3$ broader, peak exists in the SED of galaxies at far-IR wavelengths which
is caused by the integrated thermal emission from dust of different temperatures. This SED
is often parameterized as a blackbody distribution of a particular far-infrared color temperature (T$_{\rm{dust}}$)
with an additional term to account for the dust emissivity\footnote{In this paper, we assume a dust emissivity, $\beta=1.5$. The choice of emissivity affects the derived color temperature.} (e.g.~Blain et al.~2002). 

The challenge with using the far-IR bump as a redshift indicator is that the average dust temperature and the redshift are degenerate (e.g.~Blain 1999; Blain, Barnard \& Chapman 2003). Wien's displacement law tells us that the wavelength of the peak of a blackbody scales with the dust temperature. Coupling this with the displacement of the peak due to redshift we find that the observed wavelength of the far-IR peak depends linearly on both the redshift and the inverse of the dust temperature: $\lambda_{\rm{obs}}^{\rm{max}}\propto(1+z)/\rm{T}_{\rm{dust}}$.

The most luminous
galaxies in the local Universe show far-infrared color temperatures that peak at $\sim$40$\,$K although there is a significant
scatter in luminosity-temperature parameter space (e.g.~Dunne et al. 2000). For a reasonable range of dust temperatures for 
ULIRGs, 
sources that peak at 500$\,\mu$m can be found anywhere between $z\sim3$--6. 
This is a very large range in redshift and without prior information about the dust temperature it is difficult to 
further constrain the redshift with the SPIRE data alone. 

\section{High redshift candidates in the ECDFS BLAST survey}

In order to test this technique on observations we use the data from the BLAST survey of ECDFS (Devlin et al.~2009). BLAST has the same three detectors as {\it Herschel}/SPIRE but with beam sizes that are twice as big; 36, 42, 60 arcsecs FWHM at 250, 350 and 500$\,\mu$m respectively (Marsden et al.~2009). 
We use the publicly released BLAST maps\footnote{http://blastexperiment.info/results.php} and the matched filter catalogs (Chapin et al.~2010). 
These matched filter catalogs do a much better job at de-blending adjacent sources and result in higher signal-to-noise ratios than previous BLAST catalogs (see Chapin et al.~2010 for further details). 
We restrict our analysis to the central 0.47$\,$deg$^{2}$ of the BLAST image where $\sigma_{\rm{inst}}<10\,$mJy at 500$\,\mu$m.
We have conservatively added the confusion noise in quadrature with the instrument noise ($\sigma_{\rm{conf}}^{2}+\sigma_{\rm{inst}}^{2}=\sigma_{\rm{tot}}^{2}$) for all analysis in this paper. 
We only consider detections if they are $>5\sigma_{\rm{inst}}$ ($>3\sigma_{\rm{tot}}$). In this central region there are 23 galaxies robustly detected at $500\,\mu$m with $S_{500}>45\,$mJy. 
We match the 500$\,\mu$m-selected galaxies to the matched filter catalogs at 250$\,\mu$m and 350$\,\mu$m; we consider any $>5\sigma_{\rm{inst}}$ detection within a radius of 60 arcsecs to be the counterpart to the 500$\,\mu$m emission.
For sources which are undetected at these other wavelengths we set upper limits on the 350 and 250$\,\mu$m fluxes from the 90\% completeness limits of the survey (Chapin et al.~2010) which is equivalent to $3\sigma_{\rm{tot}}$.

Of these 23 500$\,\mu$m-selected galaxies, 8 have $S_{500}>S_{350}$ indicating that the dust SED peaks at or near 500$\,\mu$m -- we refer to these galaxies as 500$\,\mu$m `peakers' (although their peak may be at even longer wavelengths). 
Interestingly, none of these 500$\,\mu$m peakers (detected at $>5\sigma_{\rm{inst}}$) are detected above $5\sigma_{\rm{inst}}$ at 250 and 350$\,\mu$m. 
We inspected all of these candidates by hand in the BLAST images to make sure they are not obviously blended. 
We have not attempted to correct for flux boosting (e.g.~Coppin et al.~2006) in this analysis but, since we are not interested in the absolute fluxes and only the relative colors, this is less of a concern. In addition by considering only higher signal-to-noise ratio sources we ensure that this effect is minimal. 
In order to assess how photometric uncertainties (from both confusion and instrument noise) affect the number of 500$\,\mu$m peakers identified, we ran a Monte Carlo simulation to randomly sample the fluxes of each of the 23 sources from a Gaussian distribution $S\pm\sigma_{\rm{tot}}$. We find that photometric uncertainties can vary the number of sources with $S_{500}>S_{350}$ by 2. Thus our final number of 500$\,\mu$m peakers is $8\pm2$. 

\begin{figure}
\begin{center}
\includegraphics[width=6in,angle=0]{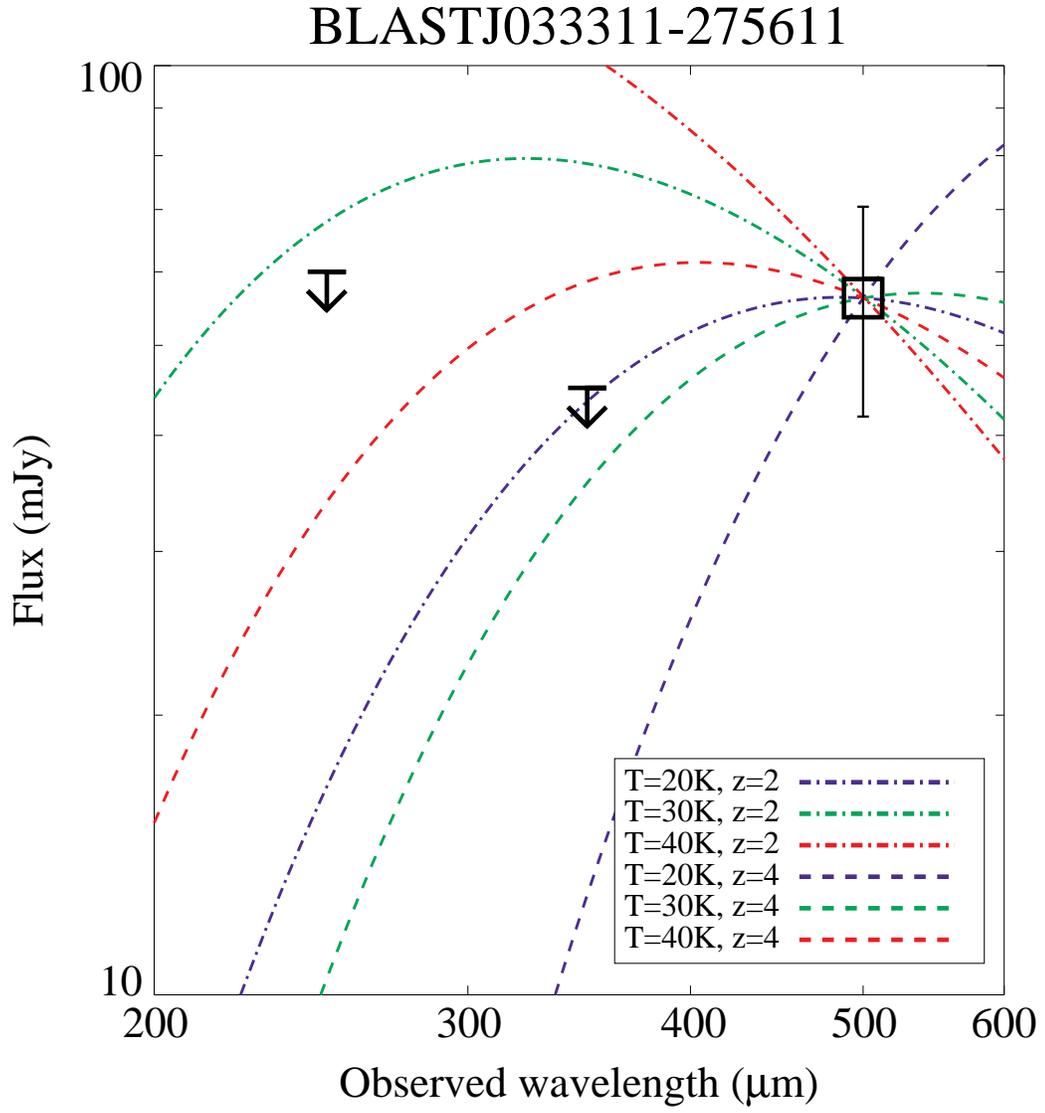}
\caption{Example far-IR SED of one of our high redshift candidates from the BLAST ECDFS survey. The current data and limits are consistent with very cold $z\sim2$ sources as well as $z\sim4$ sources with average dust temperatures.
Stronger limits at 250 and 350\,$\mu$m will be required to alleviate this degeneracy.
}
\label{fig:sed}
\end{center}
\end{figure}

These 8 high redshift candidates are listed in Table 1 along with their fluxes and Fig.~\ref{fig:sed} shows an example SED for one source. 
Five of these 500$\,\mu$m peakers are within the ECDFS field and one is within the smaller GOODS-S region (see Table 1).
The source within GOODS-S is associated (6 arcsecs away) from an AzTEC 1.1$\,$mm detected source (GS11, Scott et al.~2010); however, flux boosting in the BLAST bands prohibit a detailed comparison. 

\begin{figure}
\begin{center}
\includegraphics[width=3.in,angle=0]{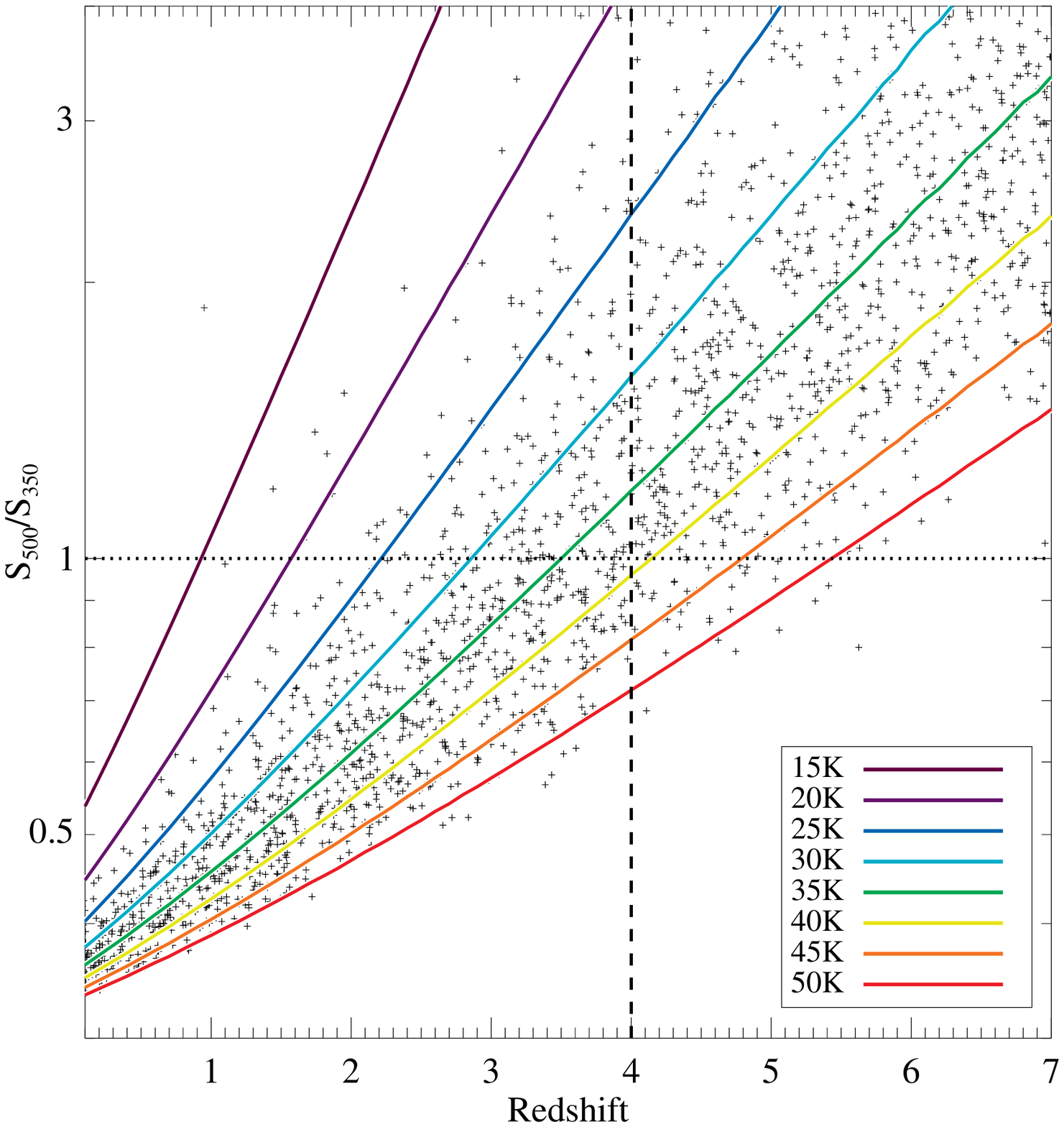}
\includegraphics[width=3.in,angle=0]{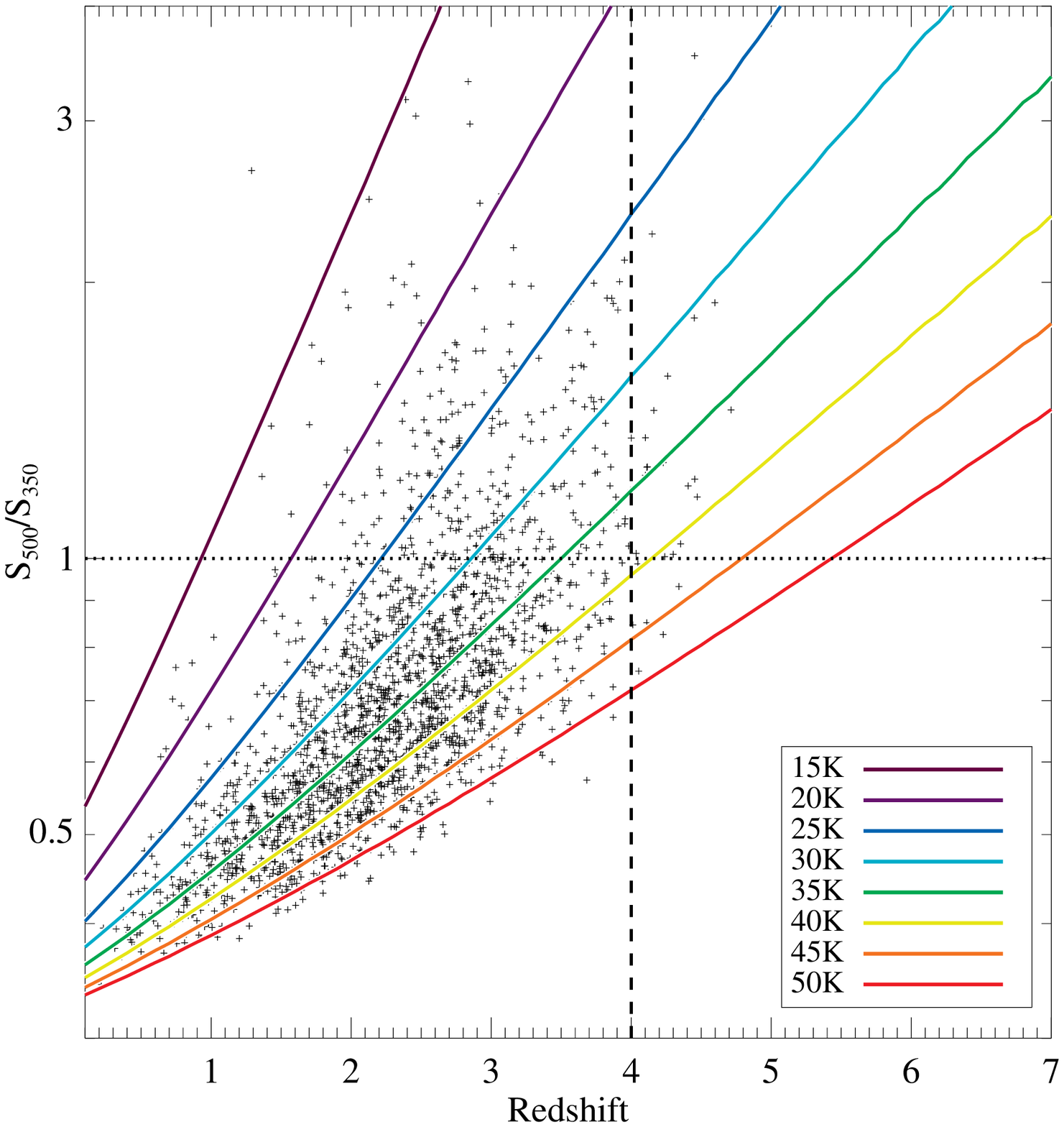}
\caption{Criteria for selecting 500$\,\mu$m peakers and the degeneracy between dust temperature and redshift. 
The curves show the S$_{500}$/S$_{350}$ flux density ratio of an SED (single temperature modified blackbody with $\beta=1.5$) for a given T$_{\rm{dust}}$ (see color legend) as a function of redshift.
The horizontal dotted line shows the criteria used to define a source that peaks at 500$\,\mu$m; there is a range of redshift and dust temperature combinations that produces this ratio. 
The data points show a Monte Carlo simulation of 2000 galaxies following a dust temperature distribution of $35+/-7\,$K, and a uniform redshift distribution (left panel) or $z=2.2+/-0.8$ (right panel).
The uniform redshift distribution results in the majority of 500$\,\mu$m peakers being at $z>4$, whereas the Gaussian redshift distribution puts very few sources above $z=4$ and the majority of 500$\,\mu$m peakers at $z<4$.
}
\label{fig:Tdust}
\end{center}
\end{figure}

Knowing that a source peaks at (or near) 500$\,\mu$m does not provide a definitive redshift due to the degeneracy between redshift and temperature (e.g.~Fig.~\ref{fig:Tdust}). 
If we conservatively assume that all 500$\,\mu$m peakers are at $z>4$ we can obtain an upper limit on the number density of $z>4$ galaxies predicted in {\it Herschel} SPIRE surveys. Based on our analysis of the BLAST data, the number density of galaxies (with S$_{500}>45\,$mJy) which peak at 500$\,\mu$m is $17\pm4\,$deg$^{-2}$. We expect that some of these 500$\,\mu$m peakers will be at $z<4$ since Fig.~\ref{fig:Tdust} shows that sources cooler than 40$\,$K can have $S_{500}>S_{350}$ but be at $z<4$. 

To get an estimate of the contamination of low redshift, cooler galaxies in the 500$\,\mu$m peaker sample we must assume a redshift and dust temperature distribution. Assuming a flat redshift distribution\footnote{A flat redshift distribution has an equal number of objects per redshift interval (constant dn/dz). This is roughly equivalent to a non-evolving luminosity function at $z>2$.} and a dust color temperature distribution of $35\pm7\,$K\footnote{This dust temperature distribution is consistent with local ULIRGs (Dunne et al.~2000) as well as $z\sim2$ SMGs (Chapman et al.~2005; Pope et al.~2006).}, we estimate that $15\,$\% of 500$\,\mu$m peakers will be $z<4$ (left panel of Fig.~\ref{fig:Tdust}).
In addition, we must also account for the contribution from warmer sources at $z>4$ but which peak at lower wavelengths ($S_{500}<S_{350}$). Assuming a sample of sources at $z>4$ again following a dust temperature distribution of $35\pm7\,$K, we find that only $10\,$\% have $S_{500}<S_{350}$. 
These two factors roughly cancel each other and we are left with a constraint on the number density of $z>4$ sources (with S$_{500}>45\,$mJy) of $<17\,$deg$^{-2}$.

The above limit on the number density of $z>4$ SMGs ambitiously assumes a flat redshift distribution. If instead we assume the observed redshift distribution of SMGs ($2.2\pm0.8$, Chapman et al.~2005; Pope et al.~2006), we find only 10\% of the 500$\,\mu$m peakers at $z>4$ (right panel of Fig.~\ref{fig:Tdust}) which brings the number density estimate down to $2\,$deg$^{-2}$. 
Based on our analysis of the BLAST data, we conclude that the number density of S$_{500}>45\,$mJy sources at $z>4$ is $<17\,$deg$^{-2}$ and could be as low as $2\,$deg$^{-2}$.

With some assumptions about the dust temperature distribution, we have shown that the submillimeter color selection, $S_{500}>S_{350}$ is a plausible way to identify samples containing the highest redshift SMGs selected at these wavelengths (albeit these samples contain contamination from lower redshift objects). 
Without further multi-wavelength data, it is impossible to tell individually which of the 8 candidates are at $z>4$, and which are just peaking at 500$\,\mu$m because they have cooler dust temperatures.
One way to test the reality of these high redshift candidates is to further constrain their SEDs with data at longer submm wavelengths such as those from the 870$\,\mu$m LABOCA survey of ECDFS (Weiss et al.~2009). In order to do a robust comparison between the LABOCA and BLAST fluxes, a correction for flux boosting needs to be applied to both surveys using the same method (e.g.~Coppin et al.~2006). This is beyond the scope of this paper but should be possible with full access to the signal and noise maps from both surveys. 
Another potential way to break the degeneracy between redshift and temperature for these candidates is with the radio and/or far-IR (e.g.~$100\,\mu$m) flux; we discuss this further in the next section.  

\section{Using radio and far-IR data to break the z-T degeneracy}

\begin{figure}
\begin{center}
\includegraphics[width=4.0in,angle=0]{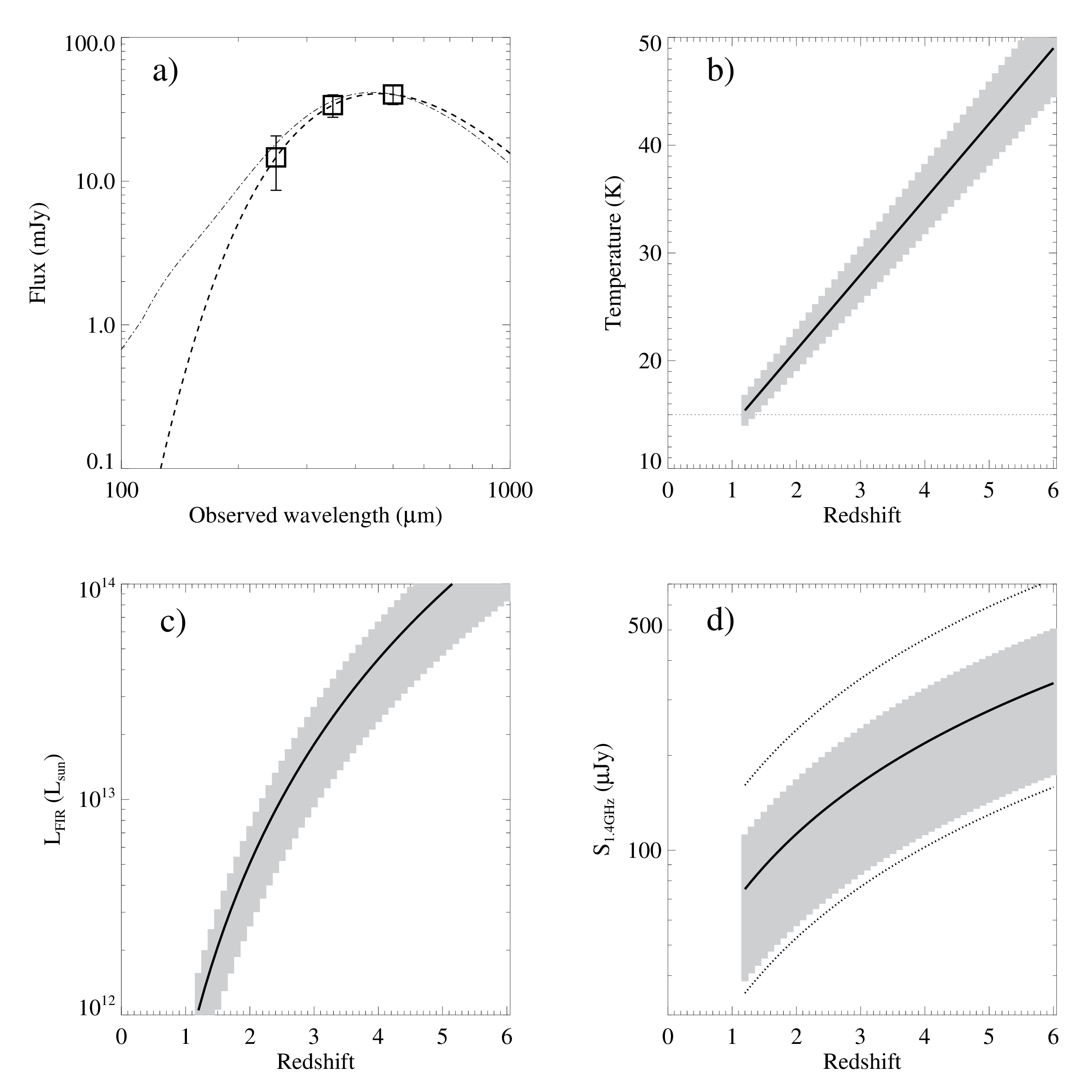}
\includegraphics[width=2.0in,angle=0]{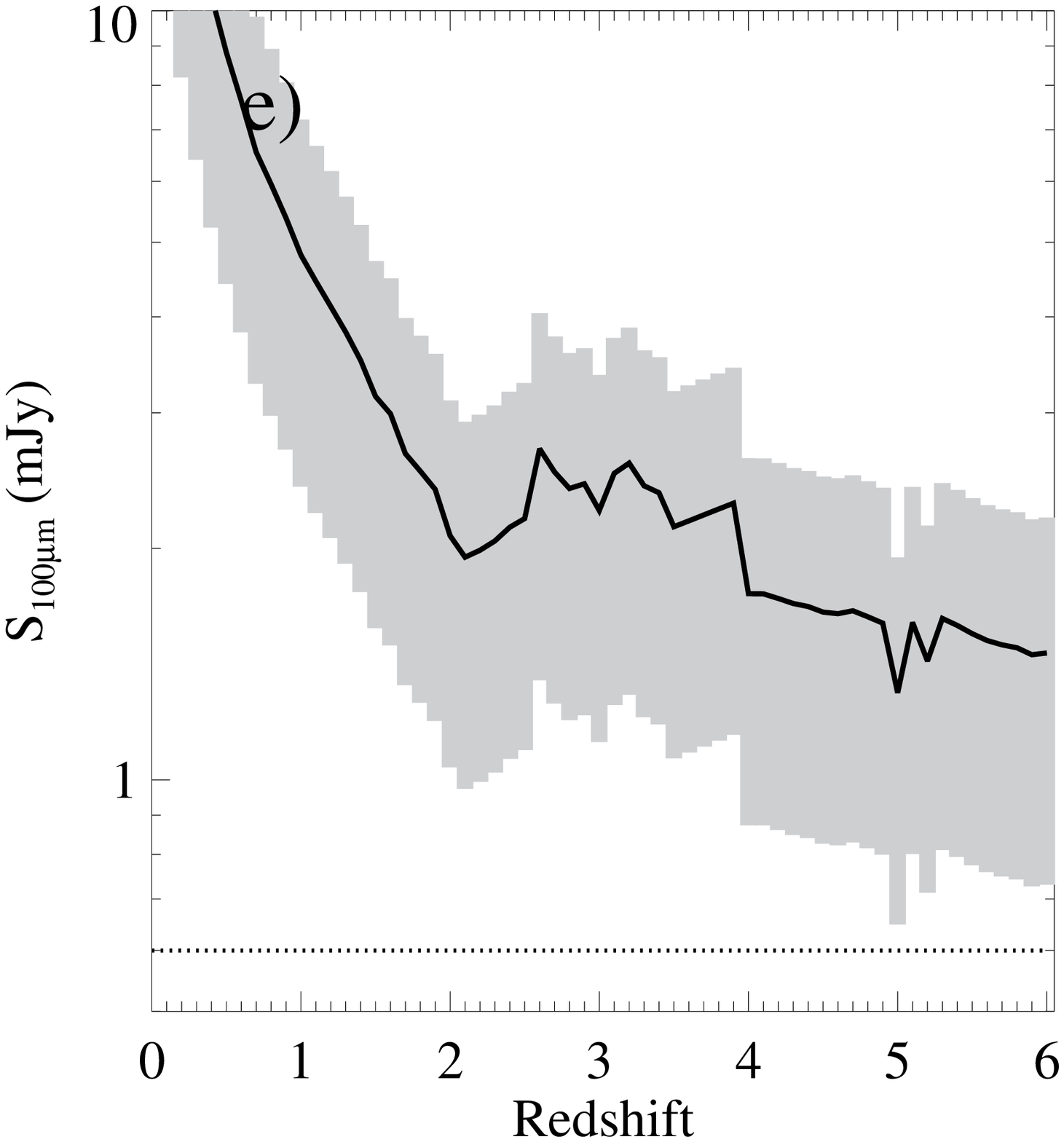}
\caption{Example of degeneracy in single temperature modified blackbody SED fitting. a) Simulated SPIRE photometry of a 500$\,\mu$m peaker (S$_{500}=40\,$mJy). Flux uncertainty includes both confusion ($\sigma_{\rm{conf}}\sim6\,$mJy) and instrument ($\sigma_{\rm{inst}}\sim1\,$mJy) noise (the shaded regions in b), c), d), and e) reflect the errors due to this measurement uncertainty). 
The dashed curve is the best-fit modified blackbody where temperature and redshift combinations can be anything along the line shown in b). We have excluded temperatures below 15K since these are unphysical due to heating by the CMB. c) The resulting $L_{\rm{FIR}}$ of the best-fit modified blackbody model as a function of redshift. d) The predicted radio flux of the best-fit modified blackbody model as a function of redshift (solid curve). The dotted curves illustrate the uncertainty resulting from the scatter in the local radio-IR correlation. The dash-dot curve in a) is a CE01 template which shows how the true SEDs of galaxies diverge from a single temperature modified blackbody at wavelengths shorter than around 50$\,\mu$m in the rest frame. e) Estimated 100$\,\mu$m flux for best-fit CE01 SED as a function of redshift. 
The dot-dash curve shows the 5$\sigma$ depth of the deepest PACS extragalactic survey, GOODS {\it Herschel} (PI D.~Elbaz). Deep radio and PACS observations can help break the degeneracy between redshift and dust temperature for the most luminous dusty galaxies.
}
\label{fig:sim}
\end{center}
\end{figure}

Photometry from 250--500$\,\mu$m alone can provide an accurate measure of $(1+z)/T_{\rm{dust}}$. Fig.~\ref{fig:sim} shows a simulation of {\it Herschel} SPIRE photometry for a source that peaks at 500$\,\mu$m. Fitting modified blackbody models to this data we find the pairs of T and z that fit these data (panel b of Fig.~\ref{fig:sim}) and the resulting $L_{\rm{FIR}}$ (panel c of Fig.~\ref{fig:sim}). 

The radio emission from galaxies is known to correlate well with the infrared emission in the local Universe (Condon 1992). Because of this radio-infrared correlation, radio data is often used to help identify counterparts to SMGs (e.g.~Ivison et al.~2002), and to constrain the redshift (e.g.~Carilli \& Yun 1999; Hughes et al.~2002; Aretxaga et al.~2007). We demonstrate the latter point in panel d of Fig.~\ref{fig:sim}. Here we assume the average value and scatter in the $q$ parameter from fitting local galaxies (Yun, Reddy \& Condon 2001). The radio flux is not degenerate with the redshift; however, unfortunately, the large scatter in the local radio-IR correlation provides fairly loose constraints on the redshift of individual objects from the radio flux. Nevertheless, a large fraction of the bright SMGs detected with SPIRE will be above the flux limits of deep ($\sigma_{\rm{radio}}<5\,\mu$Jy) 1.4 GHz radio surveys. With large statistical samples, the redshift distribution inferred from including radio observations should be robust, assuming the radio-IR correlation holds out to high redshift.

Below $\sim50\,\mu$m rest-frame, the SED of galaxies is no longer dominated by the single temperature modified blackbody but instead by warmer dust components. This can be seen in panel a of Fig.~\ref{fig:sim} where the dash-dot curve is a Chary \& Elbaz (CE01) template representative of local star forming galaxies. Fitting the CE01 templates (allowing the luminosity and temperature to vary) to the simulated SPIRE photometry we estimate the 100$\,\mu$m flux as a function of redshift (panel e of Fig.~\ref{fig:sim}). While the curve is not smooth (due to discrete templates within the CE01 library), overall the far-IR flux can constrain the redshift for this simulated galaxy with no prior assumption on its dust temperature. The dotted line in panel e of Fig.~\ref{fig:sim} shows the sensitivity of the deepest {\it Herschel} PACS extragalactic survey (GOODS {\it Herschel}, PI: D.~Elbaz); this deep survey should detect bright SMGs beyond $z\sim3$ by measuring the redshifted emission from warm dust that is heated in O\&B star-forming regions and appears to correlate
well with ongoing star-formation (Calzetti et al. 2005).

\section{Discussion}

From the BLAST ECDFS data we estimate a space density of $<17\,$deg$^{-2}$ for sources with S$_{500}>45\,$mJy at $z>4$; this is $<35\,$\% (8/23) of all sources selected down to these depths in the BLAST data. These bright 500$\,\mu$m-selected SMGs at $z>4$ contribute $<1\%$ of the total background emission at 500$\,\mu$m (Fixsen et al.~1998). 

Constrained by a recent analysis of the infrared extragalactic background light (EBL), galaxies evolution models predict the source density of galaxies with S$_{500}>27\,$mJy\footnote{In order to compare to the models we need to assume a rough correction for flux boosting of the BLAST measurements. Based on Eales et al.~(2009), we adopt an average correction of 0.6 to the BLAST fluxes which effectively brings the flux limit down to 27$\,$mJy.} at $z>4$ to be 1--5$\,$deg$^{-2}$ (Chary \& Pope 2010). The range of predicted number density comes from considering two evolutions of the models both of which fit the observed EBL. We find that our current observational constraint is consistent with both evolutionary scenarios. 
Future submm surveys with much larger telescopes such as JCMT, LMT and CCAT will reach much deeper flux limits before hitting the confusion limit. With these surveys, we can expect to detect many more dusty galaxies at $z>4$; the models in Chary \& Pope (2010) predict a number density of 40-200$\,$deg$^{-2}$ at $z>4$ for galaxies down to 5$\,$mJy at 500$\,\mu$m. 

In the past two years, the tally of $z>4$ spectroscopically confirmed SMGs has grown from zero to five\footnote{A sixth $z>4$ SMG recently appeared in the literature (Knudsen et al.~2010).} (Capak et al.~2008; Coppin et al.~2009; Daddi et al.~2009a,b). 
Coppin et al.~(2009) estimate the number density of SMGs at $z>4$ by combining these 5 sources from 3 independent submillimeter surveys and find a lower limit of $>7\,$deg$^{-2}$. While this is consistent with our predictions for 500$\,\mu$m-selected galaxies we note that the sensitivities of the BLAST and submillimeter surveys are not the same at $z=4$; only $1/5$ of the $z>4$ spectroscopically confirmed SMGs might be detected down to the BLAST ECDFS survey depths (assuming a typical SMG dust temperature). Folding this in, the Coppin et al.~(2009) estimate down to S$_{850}>20\,$mJy scales to $>1.4\,$deg$^{-2}$ at $z>4$.

\begin{figure}
\begin{center}
\includegraphics[width=6in,angle=0]{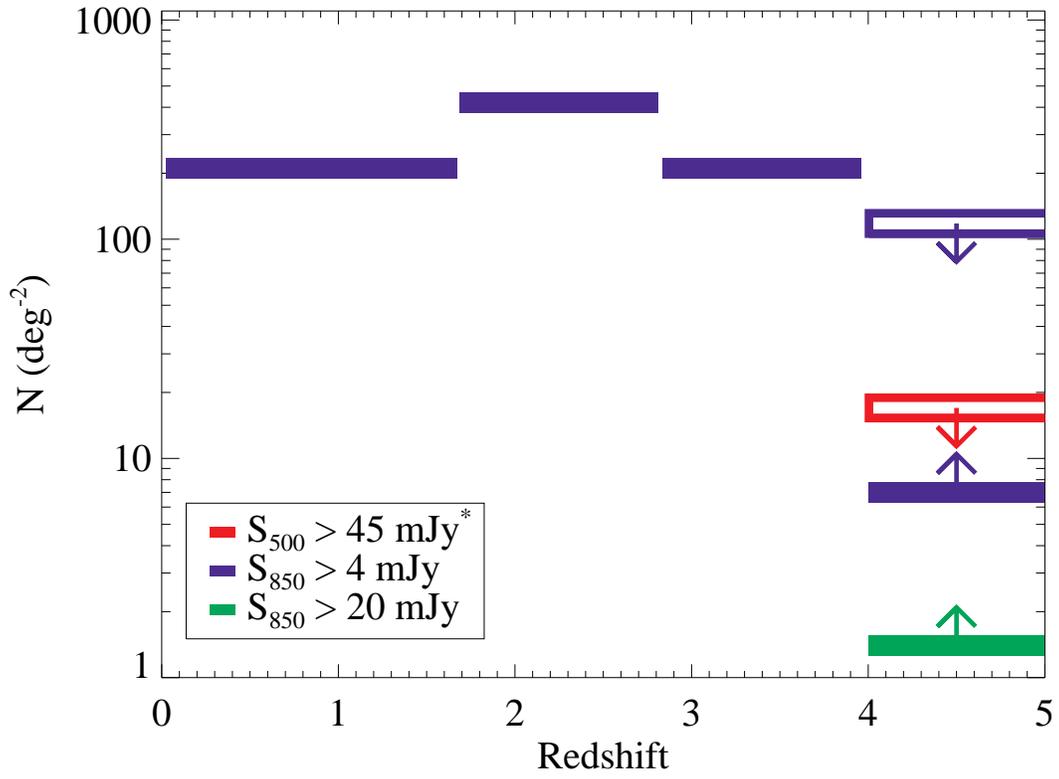}
\caption{Histogram showing the number density of submillimeter-selected galaxies as a function of redshift. 
Our new upper limit from the 500$\,\mu$m BLAST analysis is shown as the open red bar. The blue and green bars are estimates for SCUBA 850$\,\mu$m-selected galaxies from the literature where filled bars denote spectroscopic samples ($z<4$, Chapman et al.~2005; $z>4$, Coppin et al.~2009), and open bars denote photometric redshift samples (Pope et al.~2006).~$^{\ast}$Note that the BLAST sample has not been corrected for flux boosting, completeness, or spurious sources. The combination of these effects will act to decrease the true flux limit of this sample.}
\label{fig:num}
\end{center}
\end{figure}

There is some evidence that large numbers of very massive galaxies at $z>4$ pose a substantial challenge to standard galaxy formation theory (e.g.~Baugh et al.~2005). 
Fig.~\ref{fig:num} summarizes the current observational constraints on the number density of bright SMGs as a function of redshift. 
Blue and green bars are for 850$\,\mu$m-selected SMGs using values from the literature (Chapman et al.~2005; Coppin et al.~2009; Pope et al.~2006) and the red bar is our new upper limit from the analysis in this paper of the BLAST 500$\,\mu$m-selected sources. 
Overall, there does not appear to be room for a large fraction of the SMG population to be above $z=4$ down to these flux limits. 

\section{Summary}

We have shown that the degeneracy between dust temperature and redshift severely limits the use of the far-IR dust peak alone as a redshift indicator. 
The color selection of $S_{500}>S_{350}$ (and $S_{500}>S_{250}$) can be used to identify samples of candidate high redshift galaxies. In order to further constrain the redshift of individual sources requires prior knowledge of the dust temperature or additional multi-wavelength data, particularly in the radio or at 100\,$\mu$m. 

We present 8 candidate 500$\,\mu$m `peakers' from the BLAST ECDFS survey with S$_{500}>45\,$mJy; a fraction of which we expect to be at $z>4$ depending on the distribution of dust temperatures. The number density of these high redshift candidates is $<17$\,deg$^{-2}$ and is consistent with the number densities of the brightest 850$\,\mu$m-selected galaxies. The corresponding fraction of 500$\,\mu$m-selected galaxies in the BLAST survey which could be at $z>4$ is $<35\%$. 

In order to further constrain the space density of the most distant SMGs will require deep, wide area surveys with {\it Herschel}/SPIRE in addition to sufficiently deep multi-wavelength data to weed out the genuine high redshift candidates from 500$\,\mu$m `peakers' that are lower redshift
galaxies with cooler than average dust temperatures. Regardless, our upper limits on the number density of $z>4$ luminous dusty galaxies
suggests a strong decline in their number density from $z\sim2$.

\acknowledgments
We thank the referee for constructive comments which improved the presentation of this work. 
We thank the BLAST team for making their data public. 
AP acknowledges support provided by NASA through the {\it Spitzer Space Telescope} Fellowship Program, through a contract issued by the Jet Propulsion Laboratory, California Institute of Technology under a contract with NASA.

\clearpage

\begin{table}
\begin{center}
\caption{Catalog of all candidate 500$\,\mu$m `peaker' sources. IDs and fluxes come from Chapin et al.~(2010); the uncertainties listed here include both instrument and confusion noise ($\sigma_{\rm{tot}}$). The confusion noise is assumed to be 12, 13 and 15$\,$mJy at 500, 350 and 250$\,\mu$m, respectively (Chapin et al.~2010). Upper limits correspond to the flux where the catalogs are 90\% complete which is equivalent to roughly $3\sigma_{\rm{tot}}$.
The Field column indicates if the source is within the ECDFS (E) and/or the GOODS-S (G) areas.}
\vspace{0.1in}
\label{tab:all70}
\begin{tabular}{lllll}
\hline
ID & $S_{500}$ & $S_{350}$ & $S_{250}$ & Field  \\
&   (mJy) & (mJy) & (mJy)  & \\\hline
BLASTJ033311-275611	&	$56\pm14$  & $<45$  & $<60$ & E \\
BLASTJ033129-275548	&	$53\pm14$  & $<45$  & $<60$ & E  \\
BLASTJ033215-275030	&	$50\pm14$  & $<45$  & $<60$ & EG  \\
BLASTJ033254-273303	&	$50\pm14$  & $<45$  & $<60$ &   \\
BLASTJ033346-274217	&	$50\pm14$  & $<45$  & $<60$ &   \\
BLASTJ033256-280105	&	$48\pm14$  & $<45$  & $<60$ & E  \\
BLASTJ033159-273515	&	$47\pm14$  & $<45$  & $<60$ &   \\
BLASTJ033317-274926	&	$46\pm14$  & $<45$  & $<60$ & E  \\
\hline
\end{tabular}
\\
\vspace{0.1in}
\end{center}
\end{table}

\end{document}